\documentclass[twocolumn,american,aps,pre,showpacs,amsmath,amssymb]{revtex4-1}
\usepackage[latin9]{inputenc}
\usepackage{amsmath}
\usepackage{amssymb}
\usepackage{graphicx}
\usepackage{color}
\usepackage{babel}

\begin{document}

\title{Fragmentation processes in two-phase materials}

\author{H.~A.~Carmona$^{1}$}
\author{A.~V.~Guimar\~{a}es$^{1}$}
\author{J.~S.~Andrade~Jr.$^{1}$}
\author{I.~Nikolakopoulos$^{2}$}
\author{F.~K.~Wittel$^{2}$}
\author{H.~J.~Herrmann$^{1,2}$}

\affiliation{$^1$Departamento de Física, Universidade Federal do Ceará, 60451-970
Fortaleza, Ceará, Brazil}

\affiliation{$^2$Computational Physics IfB, ETH Zurich, Stefano-Franscini-Platz 3, CH-8093 Zurich,
Switzerland}

\begin{abstract}
We investigate the fragmentation process of solid materials with crystalline and
amorphous  phases using the discrete element method. Damage initiates inside
spherical samples above the contact zone in a region where the circumferential
stress field is tensile.  Cracks initiated in this region grow to form
meridional planes.  If the collision energy exceeds a critical value which
depends on the material's internal structure, cracks reach the sample surface
resulting in fragmentation.
We show that this primary fragmentation mechanism is very robust with respect
to the internal structure of the material.  For all configurations, a sharp
transition from the damage to the fragmentation regime is observed, with smaller
critical collision energies for crystalline samples.  
The mass distribution of the fragments follows a power law for small fragments
with an exponent that is characteristic for the branching merging process of
unstable cracks. Moreover this exponent depends only on the dimensionally of the
system and not on the micro structure.

\end{abstract}

\maketitle

\section{Introduction}
Multiphase materials, which are composed of different homogeneous phases, are
abundant in nature and constitute basic raw ingredients for many industrial
processes.  Comminution is an important and energy-intensive process where
various physical principles are applied to fragment multiphase material down to
a powder.  Grinding of clinker to produce the major component in Portland
cement and thus the binder in concrete~\cite{Mander:1988,Flatt2012} is an
example of a process which consumes a significant portion of the energy consumed
by mankind.  Clinker is the product of calcination of a mixture of co-grinded
minerals (80\%) and clays (20\%). The blended compound is a complex mineral
product composed of at least four principal mineral phases $C_{3}S$, $C_{2}S$,
$C_{3}A$, $C_{4}AF$, (in cement chemist notation)~\cite{Aitcin20001349}.  The
first two phases (Alite (50-65\%) and Belite (10-20\%) are crystalline, with the
rest being amorphous.  Several improvements have been conceived
~\cite{Flatt2012} in order to boost the efficiency of the comminution process.
In particular fragmentation studies have shown the potential to substantially
reduce the energy consumption of the overall cement
production~\cite{Misra1999,Cleary2008}. 

In the past, statistical models and corresponding simulation schemes have been
developed to systematically investigate brittle fragmentation
~\cite{Astrom:2006nl,Herrmann:1989nl,Kun:1999cu, Diehl:2000rd, Araripe2005,
Araripe2005a}  in terms of the resulting fragment size distribution, crack
merging and propagation, instability and branching, and the occurrence of damage
transition.  Simulations based on Lennard-Jones (MD) systems, continuum-,
elastic element-, beam- and lattice models have been able to reproduce quite
nicely the observed behavior. However for the sake of simplicity and to increase
computational efficiency, most fragmentation simulations only consider
single-phase materials.  In this work we address the brittle fragmentation
process of multiphase materials considering the simplest case of two-phase
materials, where a crystalline elastic phase is embedded in an amorphous elastic
matrix.  The non-isotropic structure of the crystalline phase is taken into
account explicitly through a hexagonal close packing (hcp) lattice. During
comminution, the material is reduced from macroscopic granules to a microscopic
powder.  We will then compare fragmentation of purely crystalline and amorphous
samples with the bi-phase ones throughout this study to explore the effect of
texture. 

This paper is organized as follows. In Sec.~\ref{sec:model_description} we
describe the model used in this work, explaining how multiple phases are
introduced in conjunction with the Discrete Element Model scheme. 
In Sec.~\ref{sec:impact_simulations} results for impact simulations for amorphous,
crystalline and two-phase materials are presented and the occurrence of
different fragmentation mechanisms, fragmentation regimes and resulting fragment
mass distributions are analysed. Finally in Sec.~\ref{sec:conclusions} we
present the conclusions and perspective of future work.

\section{Model description \label{sec:model_description}} 

The most successful numerical approaches to dynamic fragmentation so far are
based on the discrete-element model (DEM)~\cite{Cundall:1979ih}.  This type of
technique has been largely used for the simulation of ball
mills~\cite{Mishra:1992,Cleary2004,Weerasekara20133}, shear
flow~\cite{Latzel:2000,Iwashita:2000,Eggersdorfer:2010,Kamrin2012}, compaction
~\cite{Grof2012}, and fracture of
materials~\cite{Potapov:1996bz,Kun:1999cu,Wittel:2005hk,Carmona:2007yr,Timar:2010uq,Timar:2012fk,Asahina2011,Affes:2012},
among numerous other applications with particles of various shapes and diverse
cohesive elements.  The three-dimensional DEM used in this
work~\cite{Carmona:2008zg} discretizes the material by an assembly of $N^p$
spherical elements with different sizes.  For calculating a repulsive Hertzian
contact force $\vec{F}_{c}$ as an elastic interaction, a finite stiffness $E^p$
is assigned, so that two particles are allowed to overlap slightly.  Particles
are bonded by $N^b$ cylindrical beam-truss elements that may deform by
elongation, bending and torsion, producing bond forces $\vec{F}_{b}$ and moments
$\vec{M}_{b}$ on the corresponding particle centers.  A detailed description of
the force computation (normal, shear and damping forces and moments), as well as
the 3D representation of the beam-truss elements used in this work can be found
in Refs.~~\cite{Poschel:2005ec,Carmona:2008zg}.  The time evolution of the
system, namely translation and rotation of each particle, is followed
numerically by solving Newton's equation of motion through explicit numerical
integration with a time increment $\Delta t$ ~\cite{Carmona:2008zg}.  Dynamic
fracturing of the material is incorporated into the model through the sequential
failure of beam-truss elements. Beams are removed once their elliptical breaking
rule~\cite{Kun:1996rx,Kun:1996wb,Herrmann:1989nl,Carmona:2008zg} based on the
von Mises criterion type~\cite{Herrmann:1989nl}  
\begin{equation*}
    \left(\frac{\varepsilon}{\varepsilon_{th}}\right)^{2} +
    \frac{\max\left(\left|\theta_{i}\right|,\left|\theta_{j}\right|\right)}{\theta_{th}}\ge1
\end{equation*}
is fulfilled, where $\varepsilon$ is the longitudinal strain and $\theta_{i}$
and $\theta_{j}$ are the general rotation angles at the ends of the beam
connecting particle $i$ with $j$, respectively. The threshold values,
$\varepsilon_{th}$ and $\theta_{th}$, are sampled from a Weibull distribution
~\cite{Weibull:1951,Carmona:2008zg}, introducing quenched disorder in the
system. The beam breaking mechanism is irreversible in the sense that broken beams are
excluded from the force calculations for all consecutive time steps. The
macroscopic strength of the material can be tuned by adopting the average
breaking threshold and the amount of disorder in each material phase separately.

\begin{figure*} 
    \begin{tabular}{ccc} 
        \includegraphics[width=5cm]{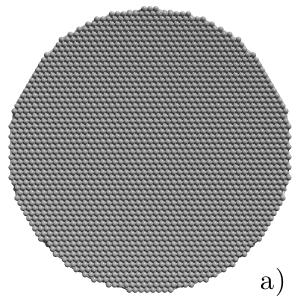} & 
        \includegraphics[width=5cm]{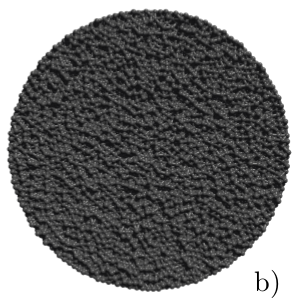} &
        \includegraphics[width=5cm]{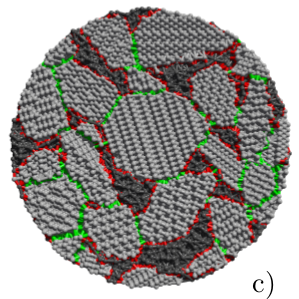} \tabularnewline
    \end{tabular} 
    \caption{ (Color online) Cross sections of crystalline (a), amorphous (b) and multiphase
    samples (c).  Colors represent different phases. In (c), red bonds represent
    amorphous-crystalline interfaces, while green bonds show the
    inter-crystalline interfaces.} \label{fig:samples} 
\end{figure*}

For comparison, here three different types of samples are defined as shown in
Fig.~\ref{fig:samples}.  In the case of type I, crystalline samples are
generated by placing all elements in a hexagonal close packing (hcp) regular
lattice. Nearest neighbors are connected by beam-truss elements. A spherical
sample (see Fig.\ref{fig:samples}(a)) is obtained by trimming all elements and
beams outside the desired spherical region.  The samples of type II are
amorphous solids generated by positioning particles randomly in a spherical
region and connecting them using a three-dimensional Voronoi tessellation (see
Fig.\ref{fig:samples}(b)).  The randomization of the initial configuration is
achieved by first placing elements on a hcp lattice, assigning initial random
velocities and letting the system evolve. A spherical, slowly shrinking
confinement is used to obtain a randomly packed spherical system.  Type III
samples are multiphase ones and their generation follows a more complicated
procedure.  Initially  a packing of random convex polyhedra of a desired size
and shape distribution is generated, as described in Appendix A.  A hcp particle
packing is inserted in every polyhedron with the respective local crystal
coordinate system and connected like in the method  to form crystallites.  In a
next step particles are placed in the interstitial spaces between polyhedra.
Random velocities are assigned to all particles and the system evolves inside a
spherical confinement again, until the newly added particles accommodate in the
regions between the crystallites. The diameter of the confining sphere is then
slowly decreased and the system is cooled by adding a small viscous force to all
elements.  The resulting system has ordered hcp crystals and random regions as
shown in Fig.\ref{fig:samples}(c).  Finally all particles are connected by beams
again using Voronoi tessellation.

Bonds connecting amorphous to crystalline particles are labeled
amorphous-crystalline interface, while those connecting different crystallites
are labeled inter-crystalline interface and may also be given different material
properties.   The final multiphase sample used in the simulations
performed here has a total of 46 crystallites embedded in an amorphous phase
matrix occupying a volume fraction 0.2. The average number of elements and bonds
in the crystallites is 806 and 6429, respectively.  

\section{Impact simulations \label{sec:impact_simulations}}

Single particle impact fragmentation against a rigid target is among the most
studied fragmentation scenarios both from the experimental and theoretical
perspective  \cite{Antonyuk:2006ow,Inaoka:2003ym,Salman:2002tz,Timar:2012fk}.  
For the sake of comparison, we limit ourselves to this case and
simulate impacts of various collision energies for the three distinct configurations.
The microscopic properties, namely the elastic properties of the elements and
bonds, as well as the bond breaking thresholds can be chosen to attain the
desired macroscopic stiffness and tensile strength of the respective phases.
In this work all beams are assigned identical elastic properties 
so that we focus mainly on the consequence of the underlying micro-structure.
Table~\ref{tab:material_props} summarizes all input values used in the
simulations.  
\begin{table}
    \caption{Microscopic material properties}\label{tab:material_props}

\begin{tabular}{|c|c|c|c|} \hline 
    \multicolumn{4}{|c|}{Beams}\tabularnewline\hline 
    stiffness          & $E^{b}$             & 6.0            & GPa\tabularnewline\hline 
    average length     & $\ell_{o}$          & 0.5/0.53/0.61  & mm\tabularnewline\hline 
    cross section diameter           & $d_{b}$             & 0.5            & mm\tabularnewline\hline 
    strain threshold   & $\varepsilon_{th}$  & 0.02           & \tabularnewline\hline 
    bending threshold  & $\theta_{th}$       & 3.5            & $^\circ$\tabularnewline\hline 
    Weibull shape parameter  & $m$       & 10                & \tabularnewline\hline 
    \multicolumn{4}{|c|}{Spherical elements}\tabularnewline\hline 
    stiffness  & $E^{p}$  & 3.0   & GPa\tabularnewline\hline 
    diameter   & $d_{e}$  & 0.5   & mm\tabularnewline\hline 
    density    & $\rho$   & 3000  & $\text{kg/m}^{3}$\tabularnewline\hline 
    \multicolumn{4}{|c}{Hard plate}\tabularnewline\hline 
    stiffness  & $E^{w}$  & 1000  & GPa\tabularnewline\hline 
    \multicolumn{4}{|c|}{Interaction}\tabularnewline\hline 
    friction coefficient  & $\mu$         & 1       & \tabularnewline\hline 
    Damping coefficient   & $\gamma_{n}$  & 0.0001  & $\text{kg/s}$\tabularnewline\hline 
    friction coefficient  & $\gamma_{t}$  & 0.0001  & $\text{kg/s}$\tabularnewline\hline 
    \multicolumn{4}{|c|}{System}\tabularnewline\hline 
    number of elements  & $N^{p}$            & $97058/81912/95271$    & \tabularnewline\hline 
    number of beams     & $N^{b}$            & $565174/564524/769201$ & \tabularnewline\hline 
    sphere diameter     & $D$                & 12/12/13.4             & mm\tabularnewline\hline 
\end{tabular}
\end{table}
\subsection{Fragmentation mechanisms}
\begin{figure*}
   \begin{tabular}{ccc}
      \includegraphics[width=5cm]{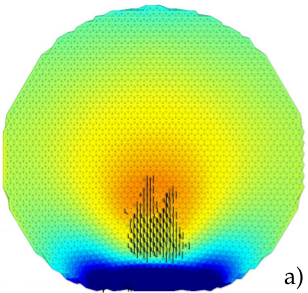}  &
      \includegraphics[width=5cm]{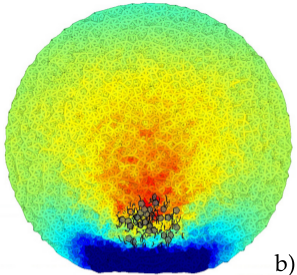}  &
      \includegraphics[width=5cm]{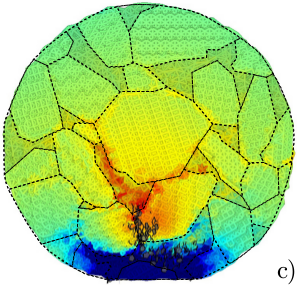} 
      \tabularnewline
   \end{tabular}
   \caption{ (Color online) Cross sections showing the damage calculated from DEM impact
   simulations for a) crystalline, b) amorphous and c) multiphase samples. Only
   beams are shown, colored according to the circumferential stress, in the
   local coordinate system, ranging from -100.0 MPa (compression) to 100 MPa
   (tension) (blue to red).  Broken bonds are represented by dark color polygons
   oriented perpendicular to their directions.} 
   \label{fig:Initial-damage}
\end{figure*}

We performed a series of numerical impact simulations of spherical samples
against a wall of stiffness $E^w \gg E^p$. As the impactor contacts the target,
it begins to deform due to repulsive contact forces. As a result, a ring of
broken bonds forms due to shear failure in the contact region.  At the same time
diffuse damage appears around this region.  It can be seen from
Fig.~\ref{fig:Initial-damage} that there is a strong correlation between the
position of the diffuse damage region and the region where the circumferential
stress in the plane perpendicular to the impact direction is tensile. In this
zone, the biaxial stress state is superimposed by a compression in impact
direction. This mechanism was reported both experimentally and numerically for
single phase materials ~\cite{Andrews:1998dg, Carmona:2008zg}.  As the
fracture evolves, cracks initiated in the biaxial stress state region develop to
form meridional cracks. If the collision energy is large enough, these cracks
propagate through the material forming crack planes that reach the sample
surface resulting in fragmentation.  Although for multiphase samples the stress
field is more heterogeneous, due to the long range correlated disorder imposed
by the different crystallites, the crack formation mechanism described above is
quite robust, resulting in meridional cracks in all three types of samples. 

The morphology of the cracks depends strongly on the texture. For crystalline
samples, cracks propagate along well defined cleavage planes of the hcp lattice.
In the amorphous samples there are no preferential orientations, but still
cracks  form meridional crack planes, cutting the sample into wedge-shaped
fragments~\cite{Carmona:2008zg}.  For multiphase samples with the more
heterogeneous stress field, meridional cracks still propagate from the biaxial
stress state region to the sample surface leading to the fragmentation of the
sample.  These cracks cut through cleavage planes in the crystalline particles
of the multiphase sample, and typically along fixed directions through the
amorphous phase.  This results in a more complicated crack morphology.

\subsection{Fragmentation regimes}

\begin{figure}
   \includegraphics[width=8cm]{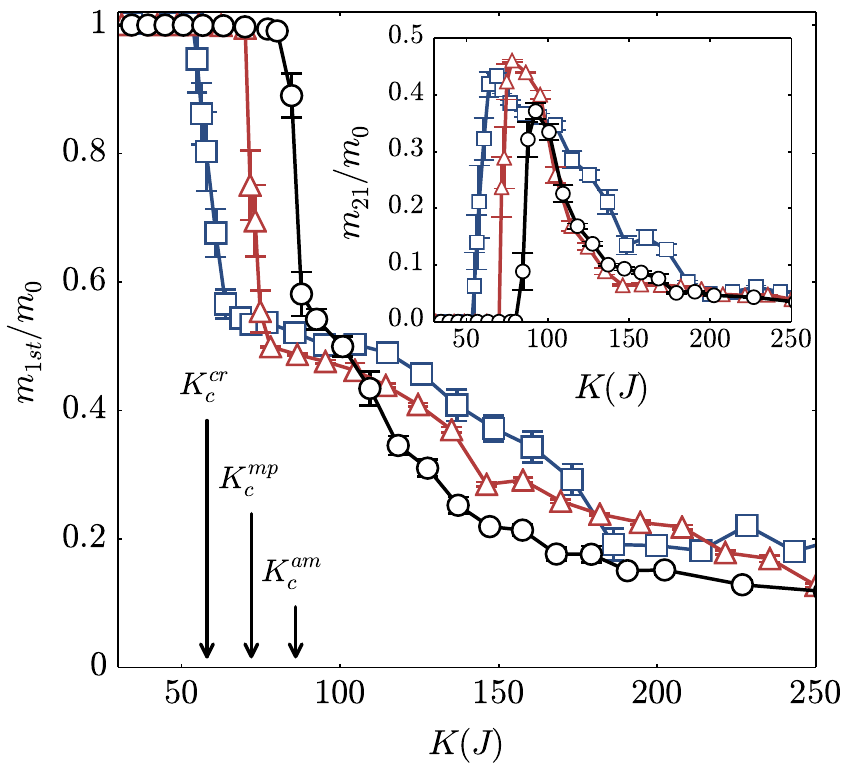}
   \caption{(Color online) Dependence of the mass of the largest fragment on the collision
   energy.  The inset shows the average total mass of all fragments excluding
   the largest. The blue square symbols correspond to crystalline samples, red
   triangles to multi-phase samples and black circles to amorphous samples.} 
   \label{fig:Collision-energy-dependence}
\end{figure}

Depending on the collision energy, impact not necessarily results in
fragmentation.  We average over 30 realizations for each energy, where fracture
thresholds of individual runs are randomly sampled from the Weibull distribution
\cite{Carmona:2008zg}.  The sizes of the final fragments depend on the collision
energy and the internal structure of the sample.  In
Fig.~\ref{fig:Collision-energy-dependence} the ensemble average of the mass of
the largest fragment, $m_{1st}$, normalized by the initial mass of the system is
plotted as a function of the collision energy $K$, for the three different
microstructures. The mass averaged of the other fragments, defined by
$m_{21}=m_{2}/m_{1}$, where $m_{k}=\sum_{i}^{N_{f}}m_{i}^{k}-m_{1st}^{k}$ with
$N_f$ being the total number of fragments,  is also shown as a function of $K$
in the inset of Fig.~\ref{fig:Collision-energy-dependence}.  
It is evident that in all three cases a collision energy exists, below which the
mass of the largest fragment corresponds to nearly the mass of the whole system.
This characterizes the damage regime as opposed to the fragmentation regime,
where the mass of the largest fragment is less than half the sample mass.  The
transition from the damage regime to the fragmentation regime occurs in a narrow
energy interval,  in which a fraction of the samples fragment, while in the
remaining only damage occurs.  We observe that in the narrow transition interval
of energy the damaged samples usually present a large crack. However,
fragmentation is prevented because the collision energy is not enough for this
crack to reach the sample's surface.  For individual samples in the ensemble,
the transition from damage to fragmentation occur more abruptly

We define the critical collision energy $K_{c}$ for each type of sample as the
energy at which the variation of the ensemble averaged mass of the largest
fragment is a maximum. The critical collision energy for the crystalline samples
is found to be $K_{c}^{cr}=58\pm4$~J.  However this value depends strongly on
the orientation of the sample lattice with respect to the impact direction. Note
that, in the results reported here, a cleavage plane contains the impact
velocity vector. For multiphase samples $K_{c}^{mp}=72\pm2$~J, this value also
depends on the sample orientation, being smaller when there are crystalline
grains in the region where cracks originate having cleave planes with normal
perpendicular to the impact velocity vector. For amorphous samples the critical
energy has the highest value of $K_{c}^{am}=86\pm2$~J.

\begin{figure}
   \includegraphics[width=8cm]{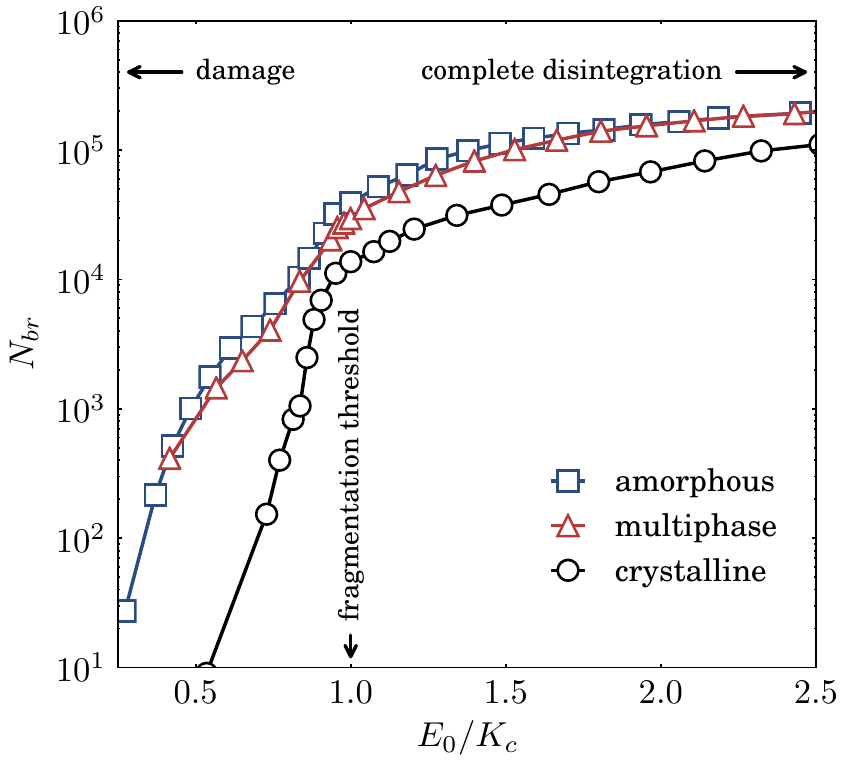}
   \caption{(Color online) Total number of broken bonds as a function of the collision energy
   normalized by $K_{c}$.}
   \label{fig:Total-broken}
\end{figure}

At $K_{c}$ the mass of largest fragment and the average mass of all other
fragments are each approximately half the initial mass of the system, indicating
that, at this energy value, one of the meridional cracks reaches the surface of the
sample cutting it into two large fragments and a few smaller ones.  Above the
critical energy, the sample disintegrates into smaller fragments.  We can observe
that the mass of the largest fragment and the average mass $m_{21}$ decay slower
as a function of $K$ for the case of crystalline and multiphase samples as
compared to the amorphous ones. 
This is because the lattice anisotropy prevents cracks from propagating in the
preferred direction, consequently hindering further fragmentation of larger
fragments by secondary mechanisms. 
The smaller critical energy for crystalline samples is also a consequence of the
anisotropy of the crystalline structure which favors the growth along well
defined cleavage planes.  This effect can be clearly seen in
Fig.~\ref{fig:Total-broken} when comparing the number of broken bonds at the
critical collision energy and hence the total dissipated energy.  As can be
expected, multiphase samples also show this effect with crack growth along
cleavage planes in crystallites, but to a much smaller extent.

\begin{figure*} 
    \begin{tabular}{ccc}
        \includegraphics[height=4.5cm]{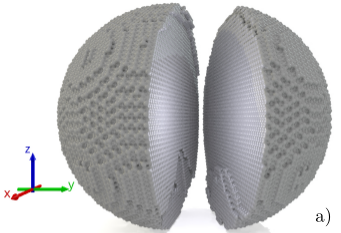}  &       
        \hspace*{-0.2cm}\includegraphics[height=4.5cm]{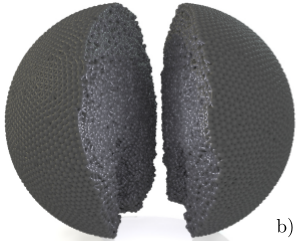}  &       
        \hspace*{-0.15cm}\includegraphics[height=4.5cm]{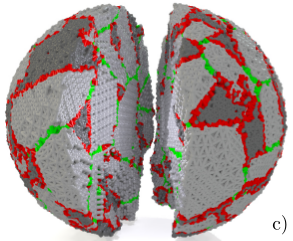}\tabularnewline
    \end{tabular}
    \caption{(Color online) The two largest fragments for: 
    a) a crystalline sample at collision energy $69\, J$ , 
    b) an amorphous sample at collision energy $94\, J$, and 
    c) a multiphase sample at collision energy $78\, J$. The fragments
    have been translated in the $y$-direction and rotated around $z$-axis for better visualization. 
    }
    \label{fig:largest_fragments}
\end{figure*}

\begin{figure}
   \begin{tabular}{ccc}
      \includegraphics[height=3.4cm]{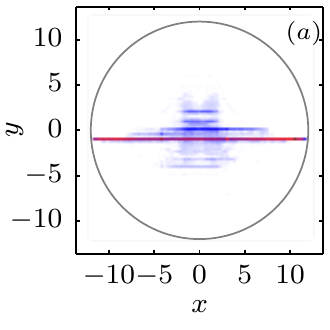}  &       
      \hspace*{-0.2cm}\includegraphics[height=3.4cm]{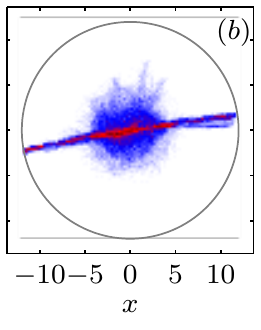}  &       
      \hspace*{-0.15cm}\includegraphics[height=3.4cm]{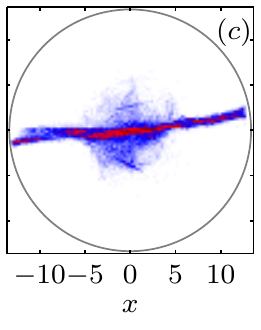}\tabularnewline
   \end{tabular}
   \caption{(Color online) Probability for a broken bond at position (x,y), perpendicular to
   the impact direction for a typical realization with: a) a crystalline sample
   at collision energy $69\, J$ , b) an amorphous sample at collision energy
   $94\, J$, and c) a multiphase sample at collision energy $78\, J$.  Colors
   correspond to the probability of having a broken bond.}
   \label{fig:Hist-position-Broken}
\end{figure}

The orientation of the resulting crack planes at the critical energy is explored
further in Figs.~\ref{fig:largest_fragments}~and~\ref{fig:Hist-position-Broken}.
Figure~\ref{fig:largest_fragments} pictures the two largest fragments resulting
from typical impact simulations with the critical energy for each of the three
types of material.  Figure~\ref{fig:Hist-position-Broken} shows the
corresponding two-dimensional histograms of the number of broken bonds for the
same simulations.  All three samples show uncorrelated bond breaking close to
the origin that corresponds to the damage in the biaxial stress zone at the
beginning of the fragmentation process.  At the final stage, well defined
diametrical planes are observed for all configurations, however the one of the
crystalline sample is the sharpest one, corresponding to a cleavage plane of the
hcp lattice.  We see that, even for the multiphase microstructure, the final
crack grows along a well defined diametrical plane.  Once a crack is formed in
the diffuse region at the beginning of the fragmentation process, it does not
change its direction until reaching the surface.  Note that the amorphous sample
exhibits more uncorrelated cracks near the impact axis than the others.
As the collision energy is increased, more meridional cracks are formed.
Azimuthal cracks, namely, cracks perpendicular to the impact axis, also appear
breaking fragments even further, in what constitutes a secondary fragmentation
mechanism.  These cracks start in a thin region, where the stress in the
direction of the impact axis is tensile due to bending of the wedge shaped
fragments, and concentrate near the contact disk.  Oblique cracks also appear
due to complex stress state that originates when the particle is already broken
into wedge-shaped fragments.

\subsection{Fragment mass distribution}
\begin{figure}
   \includegraphics[width=8cm]{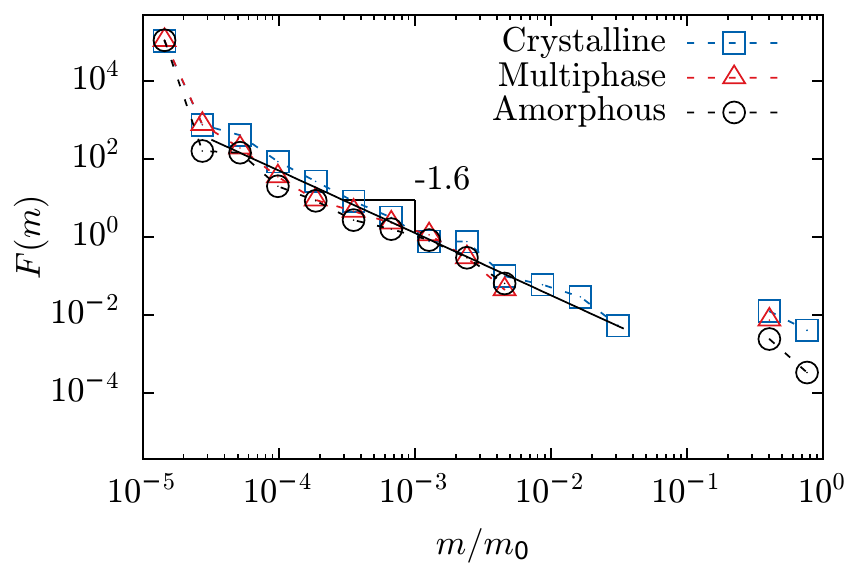}
   \caption{(Color online) Mass distribution of the fragments at the critical energy for all
   three samples.}
   \label{fig:Mass-distribution_vc}
\end{figure}

\begin{figure}
   \includegraphics[width=8cm]{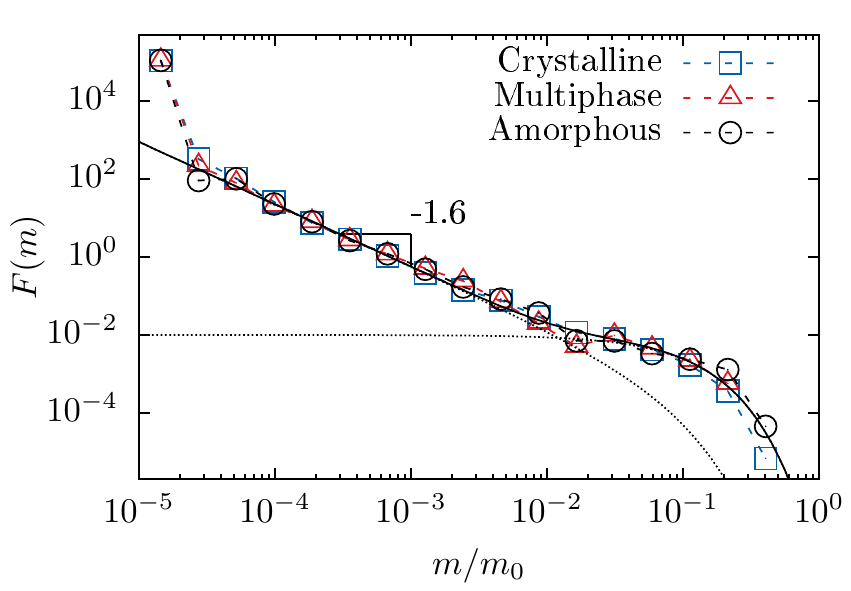}
   \caption{(Color online) Mass distribution of the fragments for $K=194$ J for all three
   samples. The solid line corresponds to the fitting using
   Eq.\eqref{eq:mdist_func}, and the dotted lines are the contributions from
   each term of this equation.}
   \label{fig:Mass-distribution-vhigh}
\end{figure}

For a more detailed analysis, the fragment mass distribution $F\left(m\right)$
at the critical energy is plotted in Fig.~\ref{fig:Mass-distribution_vc}.  Again
values are averaged over 30 realizations for each type of sample and each
collision energy.  The fragment mass distribution is surprisingly similar for
all three microstructures.  At the critical collision energy, $F\left(m\right)$
shows a peak for large fragments at about half the initial system mass.  This
corresponds to the sample breaking into two large fragments as described above.
For fragments with typically less than one percent of the system mass,
$F\left(m\right)$ follows a power law, $F\left(m\right)\sim m^{-\tau}$, with
exponent $\tau=1.6\pm0.1$.
This power law extends for $K>K_{c}$, as can be observed in
Fig.~\ref{fig:Mass-distribution-vhigh}, where the fragment mass distributions
for $K=194$~J are plotted for the three types of samples.  Surprisingly, at this
high collision energy value, the obtained mass distributions of fragments are
very similar for the three types of samples. This result suggests that, at this
point, the energy is so distant from the critical collision energy that the
particular fragmentation mechanisms causing the differences in $K_{c}$ for each
type of sample are not so relevant.  At this high collision energy, the fragment
mass distribution exhibits more clearly the power-law behavior for small
fragments.

As we can see from
Figs.~\ref{fig:Mass-distribution_vc}~and~\ref{fig:Mass-distribution-vhigh}, the
fragment mass distribution is independent of the internal material structure
within our statistical errors and can be described at high collision energy by
the expression, 
\begin{equation}
     F\left(m\right)\sim 
    (1-\beta)m^{-\tau}\exp\left(-m/\overline{m}_{0}\right)+  \beta\exp\left(-m/\overline{m}_{1}\right).\label{eq:mdist_func}
\end{equation}
This functional form has been proposed by {\AA}str\"{o}m \emph{et al}
~\cite{Astrom:2004pf,Astrom:2006nl} and has been successfully applied to
describe results both experimental and numerical results.  

The first term in Eq.\eqref{eq:mdist_func} is related to the branching and
merging process of unstable cracks. The second term originates from the
Poissonian nucleation process of dominating cracks, in our case, the meridional
cracks nucleated in the beginning of the fragmentation process.

The exponent $\tau$ depends only on the
dimensionality of the system, $\tau=(2D-1)/D$, and the parameter $\beta$
controls the relative importance of the two mechanisms. In
Fig.~\ref{fig:Mass-distribution-vhigh} Eq.~\eqref{eq:mdist_func} is plotted
using $\overline{m}_{0}=0.03\pm0.02$, $\overline{m}_{1}=0.06\pm0.06$ and
$\beta=0.9992\pm0.0005$.

\section{Conclusions \label{sec:conclusions}}

To reveal the role of the internal microstructure in the fragmentation process,
we compare impact fragmentation of spheres made of pure phases with multiphase
ones.  We employed 3D beam-truss cohesion elements with identical elastic
properties for all phases.  A transition from a damaged state to a fragmented
state is observed as the collision energy is increased.  Crystalline samples
tend to fragment at a smaller collision energy if there is a cleaveage plane that
contains the impact direction.  In this case, the dominant fracture crack
corresponds to a cleavage plane of the crystal.  For multiphase material, the
dominant crack cleaves the crystallites and cuts through the embedding amorphous
phase without changing direction. The amorphous samples require the largest
fragmentation energy.

We found that the dominant fragmentation mechanism is related to cracks that
form inside the material due to tensile radial and circumferential stress in the
ring-shaped region above the contact plane. These cracks grow to give rise to
meridional fracture planes that result in a small number of large fragments.
Even though the stress distribution is more inhomogeneous in the multiphase
material, this dominant fragmentation mechanism was found to be independent of
the internal structure of the material. As a result, the final mass distribution
of the fragments is independent of the material structure. It presents a
power-law regime for small fragments and a broad exponential region for large
fragments.  The fragment mass distribution can be successfully explained in
terms of the branching and merging processes of unstable cracks and the
Poissonian initiation process for the dominant cracks. 

The influence on fragmentation of the size and shape dispersion of the
crystalline particles, as well as the importance of the elastic properties of
each phase, in multiphase materials give rise to interesting questions. The
ability of the model to reproduce the complex stress state and crack planes with
well defined cleavage planes in the crystalline regions opens up the possibility
to study further crack propagation problems in multiphase materials. Extensions
to different material properties for different phases and detailed studies on
the influence of size dispersion are in progress.

\section{Acknowledgments}
We thank the Brazilian Agencies CNPq, CAPES and FUNCAP for financial support.
We acknowledge support by the Swiss Commission for Technological Innovation
under grant no. KTI 13703.1 PFFLR-IW and the European Research Council through
Grant FlowCSS No. FP7-
319968 for financial support.  I.N. acknowledges support
by the Alexander S. Onassis Public Benefit Foundation from Greece.  We are also
grateful for fruitful discussions with E. Gallucci and M. Weibel from Sika
Technology AG, as well as R. J. Flatt and R. K. Mishra.

\bibliography{biblio}

\begin{thebibliography}{44}%
\makeatletter
\providecommand \@ifxundefined [1]{%
 \@ifx{#1\undefined}
}%
\providecommand \@ifnum [1]{%
 \ifnum #1\expandafter \@firstoftwo
 \else \expandafter \@secondoftwo
 \fi
}%
\providecommand \@ifx [1]{%
 \ifx #1\expandafter \@firstoftwo
 \else \expandafter \@secondoftwo
 \fi
}%
\providecommand \natexlab [1]{#1}%
\providecommand \enquote  [1]{``#1''}%
\providecommand \bibnamefont  [1]{#1}%
\providecommand \bibfnamefont [1]{#1}%
\providecommand \citenamefont [1]{#1}%
\providecommand \href@noop [0]{\@secondoftwo}%
\providecommand \href [0]{\begingroup \@sanitize@url \@href}%
\providecommand \@href[1]{\@@startlink{#1}\@@href}%
\providecommand \@@href[1]{\endgroup#1\@@endlink}%
\providecommand \@sanitize@url [0]{\catcode `\\12\catcode `\$12\catcode
  `\&12\catcode `\#12\catcode `\^12\catcode `\_12\catcode `\%12\relax}%
\providecommand \@@startlink[1]{}%
\providecommand \@@endlink[0]{}%
\providecommand \url  [0]{\begingroup\@sanitize@url \@url }%
\providecommand \@url [1]{\endgroup\@href {#1}{\urlprefix }}%
\providecommand \urlprefix  [0]{URL }%
\providecommand \Eprint [0]{\href }%
\providecommand \doibase [0]{http://dx.doi.org/}%
\providecommand \selectlanguage [0]{\@gobble}%
\providecommand \bibinfo  [0]{\@secondoftwo}%
\providecommand \bibfield  [0]{\@secondoftwo}%
\providecommand \translation [1]{[#1]}%
\providecommand \BibitemOpen [0]{}%
\providecommand \bibitemStop [0]{}%
\providecommand \bibitemNoStop [0]{.\EOS\space}%
\providecommand \EOS [0]{\spacefactor3000\relax}%
\providecommand \BibitemShut  [1]{\csname bibitem#1\endcsname}%
\let\auto@bib@innerbib\@empty
\bibitem [{\citenamefont {Mander}\ \emph {et~al.}(1988)\citenamefont {Mander},
  \citenamefont {Priestley},\ and\ \citenamefont {Park}}]{Mander:1988}%
  \BibitemOpen
  \bibfield  {author} {\bibinfo {author} {\bibfnamefont {J.~B.}\ \bibnamefont
  {Mander}}, \bibinfo {author} {\bibfnamefont {M.~J.~N.}\ \bibnamefont
  {Priestley}}, \ and\ \bibinfo {author} {\bibfnamefont {R.}~\bibnamefont
  {Park}},\ }\href@noop {} {\bibfield  {journal} {\bibinfo  {journal} {J.
  Struct. Eng.-ASCE}\ }\textbf {\bibinfo {volume} {114}},\ \bibinfo {pages}
  {1804} (\bibinfo {year} {1988})}\BibitemShut {NoStop}%
\bibitem [{\citenamefont {Flatt}\ \emph {et~al.}(2012)\citenamefont {Flatt},
  \citenamefont {Roussel},\ and\ \citenamefont {Cheeseman}}]{Flatt2012}%
  \BibitemOpen
  \bibfield  {author} {\bibinfo {author} {\bibfnamefont {R.~J.}\ \bibnamefont
  {Flatt}}, \bibinfo {author} {\bibfnamefont {N.}~\bibnamefont {Roussel}}, \
  and\ \bibinfo {author} {\bibfnamefont {C.~R.}\ \bibnamefont {Cheeseman}},\
  }\href@noop {} {\bibfield  {journal} {\bibinfo  {journal} {J. Eur. Ceram.
  Soc.}\ }\textbf {\bibinfo {volume} {32}},\ \bibinfo {pages} {2787 } (\bibinfo
  {year} {2012})}\BibitemShut {NoStop}%
\bibitem [{\citenamefont {A{\"i}tcin}(2000)}]{Aitcin20001349}%
  \BibitemOpen
  \bibfield  {author} {\bibinfo {author} {\bibfnamefont {P.-C.}\ \bibnamefont
  {A{\"i}tcin}},\ }\href@noop {} {\bibfield  {journal} {\bibinfo  {journal}
  {Cem. Concr. Res.}\ }\textbf {\bibinfo {volume} {30}},\ \bibinfo {pages}
  {1349 } (\bibinfo {year} {2000})}\BibitemShut {NoStop}%
\bibitem [{\citenamefont {Misra}\ and\ \citenamefont
  {Cheung}(1999)}]{Misra1999}%
  \BibitemOpen
  \bibfield  {author} {\bibinfo {author} {\bibfnamefont {A.}~\bibnamefont
  {Misra}}\ and\ \bibinfo {author} {\bibfnamefont {J.}~\bibnamefont {Cheung}},\
  }\href@noop {} {\bibfield  {journal} {\bibinfo  {journal} {Powder Technol.}\
  }\textbf {\bibinfo {volume} {105}},\ \bibinfo {pages} {222 } (\bibinfo {year}
  {1999})}\BibitemShut {NoStop}%
\bibitem [{\citenamefont {Cleary}\ \emph {et~al.}(2008)\citenamefont {Cleary},
  \citenamefont {Sinnott},\ and\ \citenamefont {Morrison}}]{Cleary2008}%
  \BibitemOpen
  \bibfield  {author} {\bibinfo {author} {\bibfnamefont {P.~W.}\ \bibnamefont
  {Cleary}}, \bibinfo {author} {\bibfnamefont {M.~D.}\ \bibnamefont {Sinnott}},
  \ and\ \bibinfo {author} {\bibfnamefont {R.~D.}\ \bibnamefont {Morrison}},\
  }\href@noop {} {\bibfield  {journal} {\bibinfo  {journal} {Int. J. Numer.
  Methods Fluids}\ }\textbf {\bibinfo {volume} {58}},\ \bibinfo {pages} {319}
  (\bibinfo {year} {2008})}\BibitemShut {NoStop}%
\bibitem [{\citenamefont {{\AA}str{\"o}m}(2006)}]{Astrom:2006nl}%
  \BibitemOpen
  \bibfield  {author} {\bibinfo {author} {\bibfnamefont {J.~A.}\ \bibnamefont
  {{\AA}str{\"o}m}},\ }\href@noop {} {\bibfield  {journal} {\bibinfo  {journal}
  {Adv. Phys.}\ }\textbf {\bibinfo {volume} {55}},\ \bibinfo {pages} {247}
  (\bibinfo {year} {2006})}\BibitemShut {NoStop}%
\bibitem [{\citenamefont {Herrmann}\ \emph {et~al.}(1989)\citenamefont
  {Herrmann}, \citenamefont {Hansen},\ and\ \citenamefont
  {Roux}}]{Herrmann:1989nl}%
  \BibitemOpen
  \bibfield  {author} {\bibinfo {author} {\bibfnamefont {H.~J.}\ \bibnamefont
  {Herrmann}}, \bibinfo {author} {\bibfnamefont {A.}~\bibnamefont {Hansen}}, \
  and\ \bibinfo {author} {\bibfnamefont {S.}~\bibnamefont {Roux}},\ }\href@noop
  {} {\bibfield  {journal} {\bibinfo  {journal} {Phys. Rev. B}\ }\textbf
  {\bibinfo {volume} {39}},\ \bibinfo {pages} {637} (\bibinfo {year}
  {1989})}\BibitemShut {NoStop}%
\bibitem [{\citenamefont {Kun}\ and\ \citenamefont
  {Herrmann}(1999)}]{Kun:1999cu}%
  \BibitemOpen
  \bibfield  {author} {\bibinfo {author} {\bibfnamefont {F.}~\bibnamefont
  {Kun}}\ and\ \bibinfo {author} {\bibfnamefont {H.~J.}\ \bibnamefont
  {Herrmann}},\ }\href@noop {} {\bibfield  {journal} {\bibinfo  {journal}
  {Phys. Rev. E}\ }\textbf {\bibinfo {volume} {59}},\ \bibinfo {pages} {2623}
  (\bibinfo {year} {1999})}\BibitemShut {NoStop}%
\bibitem [{\citenamefont {Diehl}\ \emph {et~al.}(2000)\citenamefont {Diehl},
  \citenamefont {Carmona}, \citenamefont {Araripe}, \citenamefont {Andrade},\
  and\ \citenamefont {Farias}}]{Diehl:2000rd}%
  \BibitemOpen
  \bibfield  {author} {\bibinfo {author} {\bibfnamefont {A.}~\bibnamefont
  {Diehl}}, \bibinfo {author} {\bibfnamefont {H.~A.}\ \bibnamefont {Carmona}},
  \bibinfo {author} {\bibfnamefont {L.~E.}\ \bibnamefont {Araripe}}, \bibinfo
  {author} {\bibfnamefont {J.~S.}\ \bibnamefont {Andrade}}, \ and\ \bibinfo
  {author} {\bibfnamefont {G.~A.}\ \bibnamefont {Farias}},\ }\href@noop {}
  {\bibfield  {journal} {\bibinfo  {journal} {Phys. Rev. E}\ }\textbf {\bibinfo
  {volume} {62}},\ \bibinfo {pages} {4742} (\bibinfo {year}
  {2000})}\BibitemShut {NoStop}%
\bibitem [{\citenamefont {Araripe}\ \emph
  {et~al.}(2005{\natexlab{a}})\citenamefont {Araripe}, \citenamefont
  {Andrade},\ and\ \citenamefont {Costa~Filho}}]{Araripe2005}%
  \BibitemOpen
  \bibfield  {author} {\bibinfo {author} {\bibfnamefont {L.~E.}\ \bibnamefont
  {Araripe}}, \bibinfo {author} {\bibfnamefont {J.~S.}\ \bibnamefont
  {Andrade}}, \ and\ \bibinfo {author} {\bibfnamefont {R.~N.}\ \bibnamefont
  {Costa~Filho}},\ }\href@noop {} {\bibfield  {journal} {\bibinfo  {journal}
  {Phys. Rev. E}\ }\textbf {\bibinfo {volume} {71}},\ \bibinfo {pages} {036119}
  (\bibinfo {year} {2005}{\natexlab{a}})}\BibitemShut {NoStop}%
\bibitem [{\citenamefont {Araripe}\ \emph
  {et~al.}(2005{\natexlab{b}})\citenamefont {Araripe}, \citenamefont {Diehl},
  \citenamefont {Andrade},\ and\ \citenamefont {Costa}}]{Araripe2005a}%
  \BibitemOpen
  \bibfield  {author} {\bibinfo {author} {\bibfnamefont {L.~E.}\ \bibnamefont
  {Araripe}}, \bibinfo {author} {\bibfnamefont {A.}~\bibnamefont {Diehl}},
  \bibinfo {author} {\bibfnamefont {J.~S.}\ \bibnamefont {Andrade}}, \ and\
  \bibinfo {author} {\bibfnamefont {R.~N.}\ \bibnamefont {Costa}},\ }\href@noop
  {} {\bibfield  {journal} {\bibinfo  {journal} {Int. J. Mod. Phys. C}\
  }\textbf {\bibinfo {volume} {16}},\ \bibinfo {pages} {253} (\bibinfo {year}
  {2005}{\natexlab{b}})}\BibitemShut {NoStop}%
\bibitem [{\citenamefont {Cundall}\ and\ \citenamefont
  {Strack}(1979)}]{Cundall:1979ih}%
  \BibitemOpen
  \bibfield  {author} {\bibinfo {author} {\bibfnamefont {P.~A.}\ \bibnamefont
  {Cundall}}\ and\ \bibinfo {author} {\bibfnamefont {O.~D.~L.}\ \bibnamefont
  {Strack}},\ }\href@noop {} {\bibfield  {journal} {\bibinfo  {journal}
  {Geotechnique}\ }\textbf {\bibinfo {volume} {29}},\ \bibinfo {pages} {47}
  (\bibinfo {year} {1979})}\BibitemShut {NoStop}%
\bibitem [{\citenamefont {Mishra}\ and\ \citenamefont
  {Rajamani}(1992)}]{Mishra:1992}%
  \BibitemOpen
  \bibfield  {author} {\bibinfo {author} {\bibfnamefont {B.}~\bibnamefont
  {Mishra}}\ and\ \bibinfo {author} {\bibfnamefont {R.~K.}\ \bibnamefont
  {Rajamani}},\ }\href@noop {} {\bibfield  {journal} {\bibinfo  {journal}
  {Applied. Math. Model.}\ }\textbf {\bibinfo {volume} {16}},\ \bibinfo {pages}
  {598 } (\bibinfo {year} {1992})}\BibitemShut {NoStop}%
\bibitem [{\citenamefont {Cleary}(2004)}]{Cleary2004}%
  \BibitemOpen
  \bibfield  {author} {\bibinfo {author} {\bibfnamefont {P.~W.}\ \bibnamefont
  {Cleary}},\ }\href@noop {} {\bibfield  {journal} {\bibinfo  {journal} {Eng.
  Computation}\ }\textbf {\bibinfo {volume} {21}},\ \bibinfo {pages} {169}
  (\bibinfo {year} {2004})}\BibitemShut {NoStop}%
\bibitem [{\citenamefont {Weerasekara}\ \emph {et~al.}(2013)\citenamefont
  {Weerasekara}, \citenamefont {Powell}, \citenamefont {Cleary}, \citenamefont
  {Evertsson}, \citenamefont {Morrison}, \citenamefont {Quist},\ and\
  \citenamefont {Carvalho}}]{Weerasekara20133}%
  \BibitemOpen
  \bibfield  {author} {\bibinfo {author} {\bibfnamefont {N.~S.}\ \bibnamefont
  {Weerasekara}}, \bibinfo {author} {\bibfnamefont {M.~S.}\ \bibnamefont
  {Powell}}, \bibinfo {author} {\bibfnamefont {P.~W.}\ \bibnamefont {Cleary}},
  \bibinfo {author} {\bibfnamefont {L.~M. T.~M.}\ \bibnamefont {Evertsson}},
  \bibinfo {author} {\bibfnamefont {R.~D.}\ \bibnamefont {Morrison}}, \bibinfo
  {author} {\bibfnamefont {J.}~\bibnamefont {Quist}}, \ and\ \bibinfo {author}
  {\bibfnamefont {R.~M.}\ \bibnamefont {Carvalho}},\ }\href@noop {} {\bibfield
  {journal} {\bibinfo  {journal} {Powder Technol.}\ }\textbf {\bibinfo {volume}
  {248}},\ \bibinfo {pages} {3 } (\bibinfo {year} {2013})}\BibitemShut
  {NoStop}%
\bibitem [{\citenamefont {L{\"a}tzel}\ \emph {et~al.}(2000)\citenamefont
  {L{\"a}tzel}, \citenamefont {Luding},\ and\ \citenamefont
  {Herrmann}}]{Latzel:2000}%
  \BibitemOpen
  \bibfield  {author} {\bibinfo {author} {\bibfnamefont {M.}~\bibnamefont
  {L{\"a}tzel}}, \bibinfo {author} {\bibfnamefont {S.}~\bibnamefont {Luding}},
  \ and\ \bibinfo {author} {\bibfnamefont {H.~J.}\ \bibnamefont {Herrmann}},\
  }\href@noop {} {\bibfield  {journal} {\bibinfo  {journal} {Granul. Matter}\
  }\textbf {\bibinfo {volume} {2}},\ \bibinfo {pages} {123} (\bibinfo {year}
  {2000})}\BibitemShut {NoStop}%
\bibitem [{\citenamefont {Iwashita}\ and\ \citenamefont
  {Oda}(2000)}]{Iwashita:2000}%
  \BibitemOpen
  \bibfield  {author} {\bibinfo {author} {\bibfnamefont {K.}~\bibnamefont
  {Iwashita}}\ and\ \bibinfo {author} {\bibfnamefont {M.}~\bibnamefont {Oda}},\
  }\href@noop {} {\bibfield  {journal} {\bibinfo  {journal} {Powder Technol.}\
  }\textbf {\bibinfo {volume} {109}},\ \bibinfo {pages} {192 } (\bibinfo {year}
  {2000})}\BibitemShut {NoStop}%
\bibitem [{\citenamefont {Eggersdorfer}\ \emph {et~al.}(2010)\citenamefont
  {Eggersdorfer}, \citenamefont {Kadau}, \citenamefont {Herrmann},\ and\
  \citenamefont {Pratsinis}}]{Eggersdorfer:2010}%
  \BibitemOpen
  \bibfield  {author} {\bibinfo {author} {\bibfnamefont {M.~L.}\ \bibnamefont
  {Eggersdorfer}}, \bibinfo {author} {\bibfnamefont {D.}~\bibnamefont {Kadau}},
  \bibinfo {author} {\bibfnamefont {H.~J.}\ \bibnamefont {Herrmann}}, \ and\
  \bibinfo {author} {\bibfnamefont {S.~E.}\ \bibnamefont {Pratsinis}},\
  }\href@noop {} {\bibfield  {journal} {\bibinfo  {journal} {J. Colloid
  Interface Sci.}\ }\textbf {\bibinfo {volume} {342}},\ \bibinfo {pages} {261 }
  (\bibinfo {year} {2010})}\BibitemShut {NoStop}%
\bibitem [{\citenamefont {Kamrin}\ and\ \citenamefont
  {Koval}(2012)}]{Kamrin2012}%
  \BibitemOpen
  \bibfield  {author} {\bibinfo {author} {\bibfnamefont {K.}~\bibnamefont
  {Kamrin}}\ and\ \bibinfo {author} {\bibfnamefont {G.}~\bibnamefont {Koval}},\
  }\href@noop {} {\bibfield  {journal} {\bibinfo  {journal} {Phys. Rev. Lett.}\
  }\textbf {\bibinfo {volume} {108}},\ \bibinfo {pages} {178301} (\bibinfo
  {year} {2012})}\BibitemShut {NoStop}%
\bibitem [{\citenamefont {Grof}\ and\ \citenamefont
  {Stepanek}(2013)}]{Grof2012}%
  \BibitemOpen
  \bibfield  {author} {\bibinfo {author} {\bibfnamefont {Z.}~\bibnamefont
  {Grof}}\ and\ \bibinfo {author} {\bibfnamefont {F.}~\bibnamefont
  {Stepanek}},\ }\href@noop {} {\bibfield  {journal} {\bibinfo  {journal}
  {Phys. Rev. E}\ }\textbf {\bibinfo {volume} {88}},\ \bibinfo {pages} {012205}
  (\bibinfo {year} {2013})}\BibitemShut {NoStop}%
\bibitem [{\citenamefont {Potapov}\ and\ \citenamefont
  {Campbell}(1996)}]{Potapov:1996bz}%
  \BibitemOpen
  \bibfield  {author} {\bibinfo {author} {\bibfnamefont {A.~V.}\ \bibnamefont
  {Potapov}}\ and\ \bibinfo {author} {\bibfnamefont {C.~S.}\ \bibnamefont
  {Campbell}},\ }\href@noop {} {\bibfield  {journal} {\bibinfo  {journal} {Int.
  J. Mod. Phys. C}\ }\textbf {\bibinfo {volume} {7}},\ \bibinfo {pages} {717}
  (\bibinfo {year} {1996})}\BibitemShut {NoStop}%
\bibitem [{\citenamefont {Wittel}\ \emph {et~al.}(2005)\citenamefont {Wittel},
  \citenamefont {Kun}, \citenamefont {Herrmann},\ and\ \citenamefont
  {Kr{\"o}plin}}]{Wittel:2005hk}%
  \BibitemOpen
  \bibfield  {author} {\bibinfo {author} {\bibfnamefont {F.~K.}\ \bibnamefont
  {Wittel}}, \bibinfo {author} {\bibfnamefont {F.}~\bibnamefont {Kun}},
  \bibinfo {author} {\bibfnamefont {H.~J.}\ \bibnamefont {Herrmann}}, \ and\
  \bibinfo {author} {\bibfnamefont {B.~H.}\ \bibnamefont {Kr{\"o}plin}},\
  }\href@noop {} {\bibfield  {journal} {\bibinfo  {journal} {Phys. Rev. E}\
  }\textbf {\bibinfo {volume} {71}},\ \bibinfo {pages} {016108} (\bibinfo
  {year} {2005})}\BibitemShut {NoStop}%
\bibitem [{\citenamefont {Carmona}\ \emph {et~al.}(2007)\citenamefont
  {Carmona}, \citenamefont {Kun}, \citenamefont {Andrade},\ and\ \citenamefont
  {Herrmann}}]{Carmona:2007yr}%
  \BibitemOpen
  \bibfield  {author} {\bibinfo {author} {\bibfnamefont {H.~A.}\ \bibnamefont
  {Carmona}}, \bibinfo {author} {\bibfnamefont {F.}~\bibnamefont {Kun}},
  \bibinfo {author} {\bibfnamefont {J.~S.}\ \bibnamefont {Andrade}}, \ and\
  \bibinfo {author} {\bibfnamefont {H.~J.}\ \bibnamefont {Herrmann}},\
  }\href@noop {} {\bibfield  {journal} {\bibinfo  {journal} {Phys. Rev. E}\
  }\textbf {\bibinfo {volume} {75}},\ \bibinfo {pages} {046115} (\bibinfo
  {year} {2007})}\BibitemShut {NoStop}%
\bibitem [{\citenamefont {Tim{\'a}r}\ \emph {et~al.}(2010)\citenamefont
  {Tim{\'a}r}, \citenamefont {Bl{\"o}mer}, \citenamefont {Kun},\ and\
  \citenamefont {Herrmann}}]{Timar:2010uq}%
  \BibitemOpen
  \bibfield  {author} {\bibinfo {author} {\bibfnamefont {G.}~\bibnamefont
  {Tim{\'a}r}}, \bibinfo {author} {\bibfnamefont {J.}~\bibnamefont
  {Bl{\"o}mer}}, \bibinfo {author} {\bibfnamefont {F.}~\bibnamefont {Kun}}, \
  and\ \bibinfo {author} {\bibfnamefont {H.~J.}\ \bibnamefont {Herrmann}},\
  }\href@noop {} {\bibfield  {journal} {\bibinfo  {journal} {Phys. Rev. Lett.}\
  }\textbf {\bibinfo {volume} {104}},\ \bibinfo {pages} {095502} (\bibinfo
  {year} {2010})}\BibitemShut {NoStop}%
\bibitem [{\citenamefont {Tim{\'a}r}\ \emph {et~al.}(2012)\citenamefont
  {Tim{\'a}r}, \citenamefont {Kun}, \citenamefont {Carmona},\ and\
  \citenamefont {Herrmann}}]{Timar:2012fk}%
  \BibitemOpen
  \bibfield  {author} {\bibinfo {author} {\bibfnamefont {G.}~\bibnamefont
  {Tim{\'a}r}}, \bibinfo {author} {\bibfnamefont {F.}~\bibnamefont {Kun}},
  \bibinfo {author} {\bibfnamefont {H.~A.}\ \bibnamefont {Carmona}}, \ and\
  \bibinfo {author} {\bibfnamefont {H.~J.}\ \bibnamefont {Herrmann}},\
  }\href@noop {} {\bibfield  {journal} {\bibinfo  {journal} {Phys. Rev. E}\
  }\textbf {\bibinfo {volume} {86}},\ \bibinfo {pages} {016113} (\bibinfo
  {year} {2012})}\BibitemShut {NoStop}%
\bibitem [{\citenamefont {Asahina}\ and\ \citenamefont
  {Bolander}(2011)}]{Asahina2011}%
  \BibitemOpen
  \bibfield  {author} {\bibinfo {author} {\bibfnamefont {D.}~\bibnamefont
  {Asahina}}\ and\ \bibinfo {author} {\bibfnamefont {J.~E.}\ \bibnamefont
  {Bolander}},\ }\href@noop {} {\bibfield  {journal} {\bibinfo  {journal}
  {Powder Technol.}\ }\textbf {\bibinfo {volume} {213}},\ \bibinfo {pages} {92
  } (\bibinfo {year} {2011})}\BibitemShut {NoStop}%
\bibitem [{\citenamefont {Affes}\ \emph {et~al.}(2012)\citenamefont {Affes},
  \citenamefont {Delenne}, \citenamefont {Monerie}, \citenamefont {Radjai},\
  and\ \citenamefont {Topin}}]{Affes:2012}%
  \BibitemOpen
  \bibfield  {author} {\bibinfo {author} {\bibfnamefont {R.}~\bibnamefont
  {Affes}}, \bibinfo {author} {\bibfnamefont {J.~Y.}\ \bibnamefont {Delenne}},
  \bibinfo {author} {\bibfnamefont {Y.}~\bibnamefont {Monerie}}, \bibinfo
  {author} {\bibfnamefont {F.}~\bibnamefont {Radjai}}, \ and\ \bibinfo {author}
  {\bibfnamefont {V.}~\bibnamefont {Topin}},\ }\href@noop {} {\bibfield
  {journal} {\bibinfo  {journal} {Eur. Phys. J. E}\ }\textbf {\bibinfo {volume}
  {35}} (\bibinfo {year} {2012})}\BibitemShut {NoStop}%
\bibitem [{\citenamefont {Carmona}\ \emph {et~al.}(2008)\citenamefont
  {Carmona}, \citenamefont {Wittel}, \citenamefont {Kun},\ and\ \citenamefont
  {Herrmann}}]{Carmona:2008zg}%
  \BibitemOpen
  \bibfield  {author} {\bibinfo {author} {\bibfnamefont {H.~A.}\ \bibnamefont
  {Carmona}}, \bibinfo {author} {\bibfnamefont {F.~K.}\ \bibnamefont {Wittel}},
  \bibinfo {author} {\bibfnamefont {F.}~\bibnamefont {Kun}}, \ and\ \bibinfo
  {author} {\bibfnamefont {H.~J.}\ \bibnamefont {Herrmann}},\ }\href
  {http://link.aps.org/doi/10.1103/PhysRevE.77.051302} {\bibfield  {journal}
  {\bibinfo  {journal} {Phys. Rev. E}\ }\textbf {\bibinfo {volume} {77}},\
  \bibinfo {pages} {051302} (\bibinfo {year} {2008})}\BibitemShut {NoStop}%
\bibitem [{\citenamefont {P{\"o}schel}\ and\ \citenamefont
  {Schwager}(2005)}]{Poschel:2005ec}%
  \BibitemOpen
  \bibfield  {author} {\bibinfo {author} {\bibfnamefont {T.}~\bibnamefont
  {P{\"o}schel}}\ and\ \bibinfo {author} {\bibfnamefont {T.}~\bibnamefont
  {Schwager}},\ }\href@noop {} {\emph {\bibinfo {title} {Computational Granular
  Dynamics : Models and Algorithms}}}\ (\bibinfo  {publisher} {Springer-Verlag
  Berlin Heidelberg New York},\ \bibinfo {year} {2005})\BibitemShut {NoStop}%
\bibitem [{\citenamefont {Kun}\ and\ \citenamefont
  {Herrmann}(1996{\natexlab{a}})}]{Kun:1996rx}%
  \BibitemOpen
  \bibfield  {author} {\bibinfo {author} {\bibfnamefont {F.}~\bibnamefont
  {Kun}}\ and\ \bibinfo {author} {\bibfnamefont {H.~J.}\ \bibnamefont
  {Herrmann}},\ }\href@noop {} {\bibfield  {journal} {\bibinfo  {journal}
  {Comput. Meth. Appl. Mech. Eng.}\ }\textbf {\bibinfo {volume} {138}},\
  \bibinfo {pages} {3} (\bibinfo {year} {1996}{\natexlab{a}})}\BibitemShut
  {NoStop}%
\bibitem [{\citenamefont {Kun}\ and\ \citenamefont
  {Herrmann}(1996{\natexlab{b}})}]{Kun:1996wb}%
  \BibitemOpen
  \bibfield  {author} {\bibinfo {author} {\bibfnamefont {F.}~\bibnamefont
  {Kun}}\ and\ \bibinfo {author} {\bibfnamefont {H.~J.}\ \bibnamefont
  {Herrmann}},\ }\href@noop {} {\bibfield  {journal} {\bibinfo  {journal} {Int.
  J. Mod. Phys. C}\ }\textbf {\bibinfo {volume} {7}},\ \bibinfo {pages} {837}
  (\bibinfo {year} {1996}{\natexlab{b}})}\BibitemShut {NoStop}%
\bibitem [{\citenamefont {Weibull}(1951)}]{Weibull:1951}%
  \BibitemOpen
  \bibfield  {author} {\bibinfo {author} {\bibfnamefont {W.}~\bibnamefont
  {Weibull}},\ }\href@noop {} {\bibfield  {journal} {\bibinfo  {journal} {J.
  Appl. Mech. T. ASME}\ }\textbf {\bibinfo {volume} {18}},\ \bibinfo {pages}
  {293} (\bibinfo {year} {1951})}\BibitemShut {NoStop}%
\bibitem [{\citenamefont {Antonyuk}\ \emph {et~al.}(2006)\citenamefont
  {Antonyuk}, \citenamefont {Khanal}, \citenamefont {Tomas}, \citenamefont
  {Heinrich},\ and\ \citenamefont {Morl}}]{Antonyuk:2006ow}%
  \BibitemOpen
  \bibfield  {author} {\bibinfo {author} {\bibfnamefont {S.}~\bibnamefont
  {Antonyuk}}, \bibinfo {author} {\bibfnamefont {M.}~\bibnamefont {Khanal}},
  \bibinfo {author} {\bibfnamefont {J.}~\bibnamefont {Tomas}}, \bibinfo
  {author} {\bibfnamefont {S.}~\bibnamefont {Heinrich}}, \ and\ \bibinfo
  {author} {\bibfnamefont {L.}~\bibnamefont {Morl}},\ }\href@noop {} {\bibfield
   {journal} {\bibinfo  {journal} {Chem. Eng. Process.}\ }\textbf {\bibinfo
  {volume} {45}},\ \bibinfo {pages} {838} (\bibinfo {year} {2006})}\BibitemShut
  {NoStop}%
\bibitem [{\citenamefont {Inaoka}\ and\ \citenamefont
  {Ohno}(2003)}]{Inaoka:2003ym}%
  \BibitemOpen
  \bibfield  {author} {\bibinfo {author} {\bibfnamefont {H.}~\bibnamefont
  {Inaoka}}\ and\ \bibinfo {author} {\bibfnamefont {M.}~\bibnamefont {Ohno}},\
  }\href@noop {} {\bibfield  {journal} {\bibinfo  {journal} {Fractals}\
  }\textbf {\bibinfo {volume} {11}},\ \bibinfo {pages} {369} (\bibinfo {year}
  {2003})}\BibitemShut {NoStop}%
\bibitem [{\citenamefont {Salman}\ \emph {et~al.}(2002)\citenamefont {Salman},
  \citenamefont {Biggs}, \citenamefont {Fu}, \citenamefont {Angyal},
  \citenamefont {Szabo},\ and\ \citenamefont {Hounslow}}]{Salman:2002tz}%
  \BibitemOpen
  \bibfield  {author} {\bibinfo {author} {\bibfnamefont {A.~D.}\ \bibnamefont
  {Salman}}, \bibinfo {author} {\bibfnamefont {C.~A.}\ \bibnamefont {Biggs}},
  \bibinfo {author} {\bibfnamefont {J.}~\bibnamefont {Fu}}, \bibinfo {author}
  {\bibfnamefont {I.}~\bibnamefont {Angyal}}, \bibinfo {author} {\bibfnamefont
  {M.}~\bibnamefont {Szabo}}, \ and\ \bibinfo {author} {\bibfnamefont {M.~J.}\
  \bibnamefont {Hounslow}},\ }\href@noop {} {\bibfield  {journal} {\bibinfo
  {journal} {Powder Technol.}\ }\textbf {\bibinfo {volume} {128}},\ \bibinfo
  {pages} {36} (\bibinfo {year} {2002})}\BibitemShut {NoStop}%
\bibitem [{\citenamefont {Andrews}\ and\ \citenamefont
  {Kim}(1998)}]{Andrews:1998dg}%
  \BibitemOpen
  \bibfield  {author} {\bibinfo {author} {\bibfnamefont {E.~W.}\ \bibnamefont
  {Andrews}}\ and\ \bibinfo {author} {\bibfnamefont {K.~S.}\ \bibnamefont
  {Kim}},\ }\href@noop {} {\bibfield  {journal} {\bibinfo  {journal} {Mech.
  Mater.}\ }\textbf {\bibinfo {volume} {29}},\ \bibinfo {pages} {161} (\bibinfo
  {year} {1998})}\BibitemShut {NoStop}%
\bibitem [{\citenamefont {{\AA}str{\"o}m}\ \emph {et~al.}(2004)\citenamefont
  {{\AA}str{\"o}m}, \citenamefont {Linna}, \citenamefont {Timonen},
  \citenamefont {Moller},\ and\ \citenamefont {Oddershede}}]{Astrom:2004pf}%
  \BibitemOpen
  \bibfield  {author} {\bibinfo {author} {\bibfnamefont {J.~A.}\ \bibnamefont
  {{\AA}str{\"o}m}}, \bibinfo {author} {\bibfnamefont {R.~P.}\ \bibnamefont
  {Linna}}, \bibinfo {author} {\bibfnamefont {J.}~\bibnamefont {Timonen}},
  \bibinfo {author} {\bibfnamefont {P.~F.}\ \bibnamefont {Moller}}, \ and\
  \bibinfo {author} {\bibfnamefont {L.}~\bibnamefont {Oddershede}},\
  }\href@noop {} {\bibfield  {journal} {\bibinfo  {journal} {Phys. Rev. E}\
  }\textbf {\bibinfo {volume} {70}},\ \bibinfo {pages} {026104} (\bibinfo
  {year} {2004})}\BibitemShut {NoStop}%
\bibitem [{\citenamefont {Nikolakopoulos}(2013)}]{ilias2013}%
  \BibitemOpen
  \bibfield  {author} {\bibinfo {author} {\bibfnamefont {I.}~\bibnamefont
  {Nikolakopoulos}},\ }\emph {\bibinfo {title} {Clinker Simulation}},\
  \href@noop {} {Master's thesis},\ \bibinfo  {school} {Swiss Federal Institute
  of Technology (ETH Zurich)} (\bibinfo {year} {2013})\BibitemShut {NoStop}%
\bibitem [{\citenamefont {Gonzalez}\ \emph {et~al.}(2003)\citenamefont
  {Gonzalez}, \citenamefont {Woods},\ and\ \citenamefont
  {Eddins}}]{Gonzalez:2003}%
  \BibitemOpen
  \bibfield  {author} {\bibinfo {author} {\bibfnamefont {R.~C.}\ \bibnamefont
  {Gonzalez}}, \bibinfo {author} {\bibfnamefont {R.~E.}\ \bibnamefont {Woods}},
  \ and\ \bibinfo {author} {\bibfnamefont {S.~L.}\ \bibnamefont {Eddins}},\
  }\href@noop {} {\emph {\bibinfo {title} {Digital Image Processing Using
  MATLAB}}}\ (\bibinfo  {publisher} {Prentice-Hall, Inc.},\ \bibinfo {address}
  {Upper Saddle River, NJ, USA},\ \bibinfo {year} {2003})\BibitemShut {NoStop}%
\bibitem [{\citenamefont {OhserOhser}\ and\ \citenamefont
  {M{\"u}cklich}(2000)}]{Ohser_Muecklich}%
  \BibitemOpen
  \bibfield  {author} {\bibinfo {author} {\bibfnamefont {J.}~\bibnamefont
  {OhserOhser}}\ and\ \bibinfo {author} {\bibfnamefont {F.}~\bibnamefont
  {M{\"u}cklich}},\ }\href@noop {} {\emph {\bibinfo {title} {Statistical
  Analysis of Microstructures}}}\ (\bibinfo  {publisher} {John Wiley \& Sons,
  Ltd},\ \bibinfo {year} {2000})\BibitemShut {NoStop}%
\bibitem [{\citenamefont {Jagnow}\ \emph {et~al.}(2004)\citenamefont {Jagnow},
  \citenamefont {Dorsey},\ and\ \citenamefont {Rushmeier}}]{Jagnow2004}%
  \BibitemOpen
  \bibfield  {author} {\bibinfo {author} {\bibfnamefont {R.}~\bibnamefont
  {Jagnow}}, \bibinfo {author} {\bibfnamefont {J.}~\bibnamefont {Dorsey}}, \
  and\ \bibinfo {author} {\bibfnamefont {H.}~\bibnamefont {Rushmeier}},\ }\href
  {\doibase 10.1145/1015706.1015724} {\bibfield  {journal} {\bibinfo  {journal}
  {ACM Trans. Graph.}\ }\textbf {\bibinfo {volume} {23}},\ \bibinfo {pages}
  {329} (\bibinfo {year} {2004})}\BibitemShut {NoStop}%
\bibitem [{\citenamefont {Mishra}(2012)}]{Ratan}%
  \BibitemOpen
  \bibfield  {author} {\bibinfo {author} {\bibfnamefont {R.~K.}\ \bibnamefont
  {Mishra}},\ }\emph {\bibinfo {title} {Simulation of Interfaces in
  Construction Materials: Tricalcium Silicate, Gypsum, and Organic
  Modifiers}},\ \href@noop {} {Ph.D. thesis},\ \bibinfo  {school} {University
  of Akron, Polymer Engineering} (\bibinfo {year} {2012})\BibitemShut {NoStop}%
\bibitem [{\citenamefont {Legland}(2009)}]{Geom3d}%
  \BibitemOpen
  \bibfield  {author} {\bibinfo {author} {\bibfnamefont {D.}~\bibnamefont
  {Legland}},\ }\href@noop {} {\enquote {\bibinfo {title} {{geom3d}},}\
  }\bibinfo {howpublished}
  {\url{http://www.mathworks.com/matlabcentral/fileexchange/24484}} (\bibinfo
  {year} {2009})\BibitemShut {NoStop}%
\bibitem [{\citenamefont {D'Errico}(2012)}]{inhull}%
  \BibitemOpen
  \bibfield  {author} {\bibinfo {author} {\bibfnamefont {J.}~\bibnamefont
  {D'Errico}},\ }\href@noop {} {\enquote {\bibinfo {title} {Inhull},}\
  }\bibinfo {howpublished}
  {\url{http://www.mathworks.com/matlabcentral/fileexchange/10226-inhull}}
  (\bibinfo {year} {2012})\BibitemShut {NoStop}%
\end{thebibliography}%

\appendix
\section{Construction of multiphase sample} The grain size distribution of a
crystallic phase is a prerequisite for every microstructural simulation. Under
grain size the max caliper diameter of a convex grain is understood. Based on
two-dimensional micrographs \cite{ilias2013}, the grain size distribution of the two-dimensional
cross-sections of the grains can be recovered by means of a boundary-tracking
method \cite{Gonzalez:2003}. After making certain simplifying assumptions
regarding the grains' shape, stereological considerations allow one to estimate
the three-dimensional grain size distribution \cite{Ohser_Muecklich,Jagnow2004}.
The grain shape in our case is an irregular convex polyhedron, which renders
most of the relatively simple stereological techniques impractical. For this
reason, the assumption of a spherical shape is adopted and the problem reduces
to determining the size distribution of a polydisperse system of spherical
particles. More details on the computation of the 3D distribution estimate can
be found in \cite{Jagnow2004}.

The generation of the sample starts by producing a shape-pool composed of
individually evolved instances of one or several reference grain shapes.  For
simplicity only one phase is considered in the presentation, since other
crystallic phases are straightforward to incorporate.  The original shape could
be, for instance, a possible equilibrium shape of a crystallic phase, whose
Miller indices are estimated based on the directional cleavage energies
\cite{Ratan}. Grains are picked from this pool in a way that their sizes follow
the previously estimated grain size distribution and are placed into a confining
spherical volume at random positions but without overlap. After the desired
number of grains is in place, the size distribution constraint is already
satisfied and the shapes encode all the information regarding the crystal
structure of the phase. It then boils down to achieving a realistic volume
fraction for this phase while distorting its original characteristics as little
as possible. To this end, a simple packing algorithm is applied, making use of
the libraries \cite{Geom3d,inhull}. For a sufficiently large number of steps, a
grain is chosen at random and the following transformations are applied to it:
(i) random translation in the range $[0,\mathrm{d}s]$, (ii) random rotation
around \emph{each} axis, within the interval $[0,\mathrm{d}\theta]$ and (iii)
with probability close to $\mathrm{P}\!=\!0.5$, a random translation within
$[0,\mathrm{d}r]$ towards the center of the spherical confining volume (radial
translation). The move is only accepted if the new position does not lead to an
overlap, otherwise it is rejected and a new random grain is chosen.

\begin{figure}[h]
  \centering
  \includegraphics[width=0.5\textwidth]{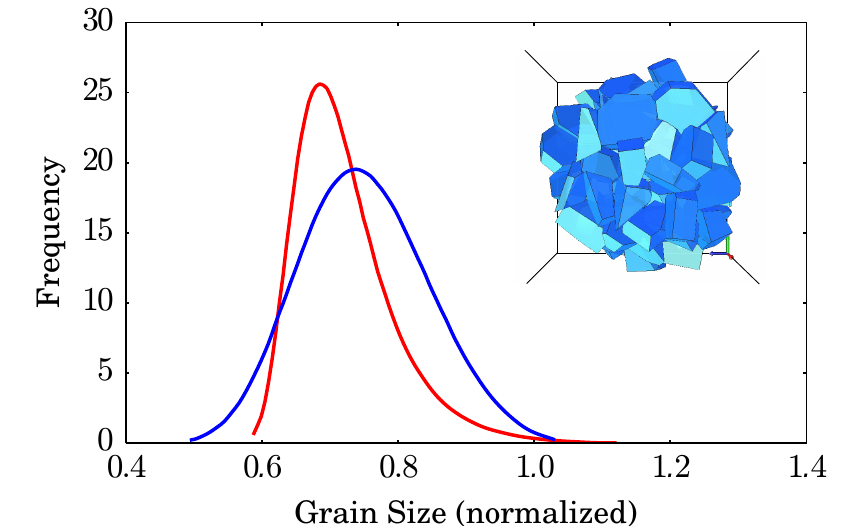}
  \caption{(Color online) Initial (red) and final (blue) grain size distributions of the
    sample. The curves depict fits to the generalized
    extreme value distribution. The grain sizes have been normalized to 1. Inset: the final
    configuration of maximal volume fraction obtained with scaling factor $1.9$.}
  \label{fig:grain_size_distributions}
\end{figure}

Since randomly shaped polyhedra are not expected to efficiently fill the space,
further action needs to be taken for a high packing density to be reached. An
expand-and-clip strategy is introduced, which targets high volume fraction at
the cost of an arbitrary (but small) shape deformation. All grains are expanded
via in-place scaling, controlled by a single scaling factor, which results in a
configuration with all grain centroids retaining their previous positions, while
arbitrary overlaps between grains occur. The non-overlap constraint is recovered
by clipping all pairs of overlapping grains with the plane that (i) has normal
parallel to the line connecting the centroids of the two involved grains and
(ii) the centroid of the convex polyhedron that forms their intersection lies on
it. Using the aforementioned procedure, a system of 100 grains was initialized
with the prescribed size distribution. After packing with $\mathrm{d}s,
\mathrm{d}r \in \left[0,0.02\right]$, $\mathrm{d}\theta
\in\left[0,\pi/100\right]$ and clipping with scaling factor $1.9$, a volume
fraction of approximately 78\% was achieved. The quality of the final sample was
determined by (i) comparison of its cross-sections with the available
micrographic data and (ii) visual inspection of the distortion of the original
grain size distribution (see Fig. \ref{fig:grain_size_distributions}).  The
sample was deemed reliable, i.e. a possible occurrence of a granular
configuration in a multiphase material.

\end{document}